%% file: main.tex
\documentclass[conference]{IEEEtran}

\usepackage[acronym]{glossaries}
\usepackage{hyperref}
\usepackage[capitalise]{cleveref}
\input{preamble.tex}

\title{Compression Robust Synthetic Speech Detection Using Patched Spectrogram Transformer}
\newcommand{\td}{$^\dagger$}
\newcommand{\tdd}{$^\ddagger$}
\author{
\parbox{0.95\linewidth}{\centering Amit Kumar Singh Yadav\td \hspace{2em} Ziyue Xiang\td  \hspace{2em} Kratika Bhagtani\td \hspace{2em}  Paolo Bestagini\tdd \\
\centering Stefano Tubaro\tdd \hspace{2em} Edward J. Delp\td
    \vspace*{0.5em}\\
    \small\centering \td  Video and Image Processing Lab (VIPER), School of Electrical and Computer Engineering, \\
    \small\centering  Purdue University, West Lafayette, Indiana, USA\\
    \small\centering \tdd Dipartimento di Elettronica, Informazione e Bioingegneria, Politecnico di Milano, Milano, Italy
}
}

\begin{document}
\maketitle
\begin{abstract}
Many deep learning  synthetic speech generation tools are readily available.
The use of synthetic speech has caused financial fraud, impersonation of people, and misinformation to spread. 
For this reason forensic methods that can detect synthetic speech have been proposed.
Existing methods often overfit on one dataset and their performance reduces substantially in practical scenarios such as detecting synthetic speech 
shared on social platforms.
In this paper we propose, \methodfullname, 
a synthetic speech detector that converts a time domain speech signal to a mel-spectrogram and processes it in patches 
using a transformer neural network.
We evaluate the detection performance of \methodname~on ASVspoof2019 dataset.
Our experiments show that \methodname~performs well on ASVspoof2019 dataset compared to other approaches using spectrogram for synthetic speech detection.
We also investigate generalization performance of \methodname~on In-the-Wild dataset.
\methodname~generalizes well than several existing methods on detecting synthetic speech from an out-of-distribution dataset.
We 
also evaluate robustness of \methodname~to detect telephone quality 
synthetic speech 
and synthetic speech shared on social platforms (compressed speech).
\methodname~is robust to compression and can detect telephone quality synthetic speech 
better than several existing methods.
\end{abstract}


\glsresetall
\input{part-1-introduction}
\input{part-2-related-work}
\input{part-3-method}
\input{part-4-results}
\input{part-5-conclusion}

\section*{Acknowledgments}
 This material is based on research sponsored by the Defense Advanced Research Projects Agency (DARPA) and the Air Force Research Laboratory (AFRL) under agreement number FA8750-20-2-1004. The U.S. Government is authorized to reproduce and distribute reprints for Governmental purposes notwithstanding any copyright notation thereon. The views and conclusions contained herein are those of the authors and should not be interpreted as necessarily representing the official policies or endorsements, either expressed or implied, of DARPA, AFRL or the U.S. Government.

Address all correspondence to Edward J. Delp, \texttt{ace@purdue.edu}.

\vspace{-.8mm}

\bibliographystyle{IEEEtran}
\bibliography{refs}

\end{document}

%% file: preamble.tex
\IEEEoverridecommandlockouts

\usepackage{cite}
\usepackage{amsmath,amssymb,amsfonts}
\usepackage{algorithmic}
\usepackage{graphicx}
\usepackage{textcomp}
\usepackage[table]{xcolor}  
\usepackage{booktabs, makecell, booktabs,multirow} 
\usepackage{url} 
\usepackage{hyperref}
\usepackage{subcaption}
\usepackage{flushend}
\usepackage{svg}

\def\BibTeX{{\rm B\kern-.05em{\sc i\kern-.025em b}\kern-.08em
    T\kern-.1667em\lower.7ex\hbox{E}\kern-.125emX}}
    
\newacronym{qmf}{QMF}{Quadrature Mirror Filter}
\newacronym{asv}{ASV}{Automatic Speaker Verification}
\newacronym{mdct}{MDCT}{Modified Discrete Cosine Transform}
\newacronym{imdct}{IMDCT}{Inverse Modified Discrete Cosine Transform}
\newacronym{mfcc}{MFCC}{Mel-Frequency Cepstrum Coefficients}
\newacronym{cqt}{CQT}{Constant-Q Transform}
\newacronym{stft}{STFT}{Short Time Fourier Transform}
\newacronym{dft}{DFT}{Discrete Fourier Transform}
\newacronym{dct}{DCT}{Discrete Cosine Transform}
\newacronym{fft}{FFT}{Fast Fourier Transform}
\newacronym{auc}{AUC}{Area Under Curve}
\newacronym{roc}{ROC}{Receiver Operating Characteristic}
\newacronym{eer}{EER}{Equal Error Rate}
\newacronym{lfccs}{LFCCs}{Linear Frequency Cepstral Coefficients}
\newacronym{mfccs}{MFCCs}{Mel Frequency Cepstral Coefficients}
\newacronym{cqccs}{CQCCs}{Constant Q Cepstral Coefficients}
\newacronym{cfccs}{CFCCs}{Cochlear Filter Cepstral Coefficients}
\newacronym{svm}{SVM}{Support Vector Machine}
\newacronym{gmm}{GMM}{Gaussian Mixture Model}
\newacronym{tssdnet}{TSSDNet}{Time-Domain Synthetic Speech  Detection Net}
\newacronym{aac}{AAC}{Advanced Audio Coding}
\newacronym{flac}{FLAC}{Free Lossless Audio Codec}
\newacronym{mlp}{MLP}{Multi Layer Perceptron Network}
\newacronym{bce}{BCE}{Binary Cross Entropy}
\newacronym{auprc}{AUPRC}{Area Under Precision Recall Curve}
\newacronym{rocauc}{ROC-AUC}{Area Under the Receiver Operating Characteristic Curve}
\newcommand{\methodname}{PS3DT}
\newcommand{\methodfullname}{Patched Spectrogram Synthetic Speech Detection Transformer (PS3DT)}
\newacronym{cnn}{CNN}{Convolutional Neural Network}
\newacronym{ssast}{SSAST}{Self-Supervised Audio Spectrogram Transformer}
\newacronym{passt}{PaSST}{Patchout faSt Spectrogram Transformer}
\newacronym{cct}{CCT}{Compact Convolution Transformer}
\newacronym{fnr}{FNR}{False Negative Rate}
\newacronym{fpr}{FPR}{False Positive Rate}
\newacronym{dnn}{DNN}{Deep Neural Network}
\newacronym{cqost}{CQ-OST}{Constant-Q Octave Subband Transform}
\newacronym{icqcc}{ICQCC}{Inverted Constant-Q Cepstral Coefficients}

\glsdisablehyper

\newcommand{\etal}{\textit{et al}.\@ }
\newcommand{\ie}{\textit{i.e.},\@ }
\newcommand{\eg}{\textit{e.g.},\@ }

\newcolumntype{L}[1]{>{\raggedright\let\newline\\\arraybackslash\hspace{0pt}}m{#1}}
\newcolumntype{C}[1]{>{\centering\let\newline\\\arraybackslash\hspace{0pt}}m{#1}}
\newcolumntype{R}[1]{>{\raggedleft\let\newline\\\arraybackslash\hspace{0pt}}m{#1}}

\definecolor{kellygreen}{rgb}{0.3, 0.73, 0.09}
\definecolor{amber}{rgb}{1.0, 0.75, 0.0}
\definecolor{amethyst}{rgb}{0.6, 0.4, 0.8}

%% file: part-1-introduction.tex
\section{Introduction}\label{part-1-intro}

By ``synthetic speech'' we mean a human sounding speech signal generated using a model rather than an actual human speaker \cite{asvspoof19, popov2021gradtts}. 
Conventional approaches for generating synthetic speech include simple waveform concatenation (WC) and source modeling (vocoders) \cite{klatt2016}.
Other approaches use neural networks~\cite{
fastspeech_2_iclr_2021, popov2021gradtts}. 
The recent use of neural networks have reduced the perceptual difference between a synthetic speech and speech from an actual human speaker (often referred to as pristine or bona fide speech)~\cite{fastspeech_2_iclr_2021, popov2021gradtts}. 
Some methods can synthesize speech impersonating any person or language accent~\cite{wellsaid,vall_e}.
These methods are useful for voice applications such as voice assistants, e-learning, movies, and advertisement.
But they have also been used to generate high quality speech for malicious purposes such as spreading misinformation~\cite{allyn_2022}, committing financial fraud, and impersonations of humans speaking~\cite{smith_2021}. 
For instance, synthetic speech was used for impersonation to commit a financial fraud worth \$40 million in 2021~\cite{smith_2021}. A deepfake video of the president of Ukraine containing synthetic speech was used to support a misinformation campaign in 2022~\cite{allyn_2022}.

There are several challenges in
developing synthetic speech detection techniques.
For example, a detector should 
work well on synthetic speech generated from unknown techniques~\cite{bhagtani2022overview}. 
Synthesized speech used to support misinformation campaigns are often shared on social media platforms such as YouTube. 
These platforms compress the speech signal using lossy compression standards such as \gls{aac}~\cite{mpeg4_book, youtube_aac}.
Hence, it is also important for a detector to be robust to compression.
Some incidents report the use of synthetic speech over telephone channels to impersonate a person or fool \gls{asv} systems.
Telephone channels distort the speech signal due to compression, packet loss, and other artifacts resulting from different bandwidths, transmission infrastructures and data rates. 
Therefore, it is also important for a detector to be robust to detect synthetic speech over telephone channels.

In this paper, we propose a transformer based approach for synthetic speech detection using mel-spectrograms. 
Existing methods using transformers for synthetic speech detection either process mel-spectrogram first using \gls{cnn} and then a transformer \cite{bartusiak_2021_asilomar}, or they process all the regions of the mel-spectrogram together \cite{bartusiak_theasis}. 
Our proposed method \methodfullname~processes the mel-spectrogram \cite{mel} in patches
and then rearranges patch representations corresponding to same temporal location together.
PS3DT shows promising detection results on ASVspoof2019 dataset compared to several existing approaches for synthetic speech detection using spectrograms \cite{rs2010}, mel-spectrograms \cite{mel} or their derivatives.
Moreover, \methodname~generalizes well compared to several existing approaches to detect synthetic speech from an out-of-distribution In-the-Wild dataset \cite{in_the_wild}.
Finally, our experiments on ASVspoof2021 dataset \cite{asvspoof_2021} show that \methodname~is more robust to compression and detecting synthetic speech over telephone than 
existing methods.
\begin{figure*}[!h]
    \centering
    \includesvg[width=\linewidth]{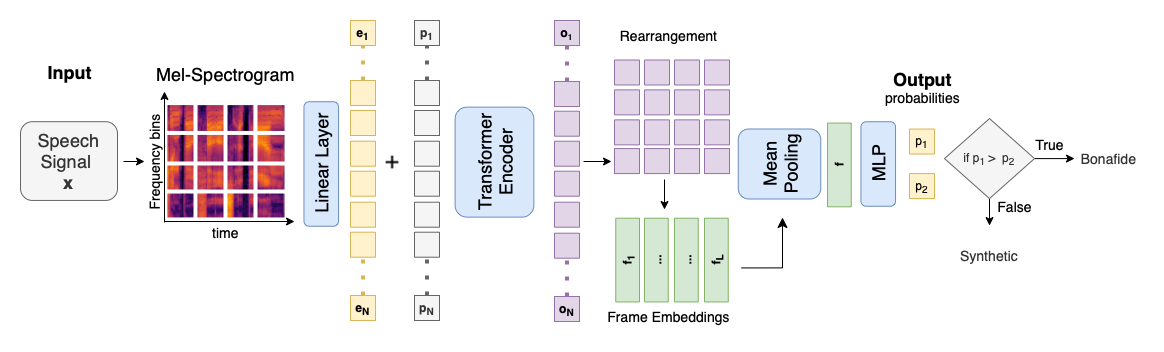}
    \caption{Block Diagram of Our Proposed: \methodfullname. }
    \label{fig:block-diagram}
\end{figure*}

%% file: part-2-related-work.tex
\section{Related Work}\label{sec:related-work}
Conventional approaches to detect synthetic speech are based 
 on classifiers using hand crafted features or time-frequency representations of the speech signal, \eg
\gls{cqt} \cite{li2021replay}, \gls{cqccs} \cite{cqccs}, \gls{mfccs} \cite{mfccs}, and \gls{lfccs} \cite{li2021replay}.
These approaches 
require tedious feature selection procedures~\cite{zakariah2017digital}. 
Other approaches are based on 
analyzing
the speech signal in the time domain. 
For example, Guang \etal~propose a neural network that uses time domain speech and processes it as a sequence of information using a recurrent neural network~\cite{he2016deep}.
They show promising results as compared to the hand-crafted feature based approaches.
Another family of approaches rely on the use of \glspl{cnn}, treating time-frequency audio representations as images. Bartusiak~\etal~\cite{bartusiak_2021_asilomar} and Subramani~\etal \cite{subramani2020learning} show that spectrograms and computer vision approaches can be used for detecting synthetic speech.
A spectrogram is a 2D representation of a speech signal~\cite{rs2010}.
The horizontal axis represents time and the vertical axis represents frequency~\cite{long-mipr, alan_acm_2023}.
If the frequency axis is in the mel scale and not in the Hertz scale, it is known as a mel-spectrogram~\cite{mel}.
The conversion between Hertz frequency scale $f_\text{Hz}$ and the mel frequency scale  $f_\text{mel}$~\cite{mel} is obtained by
\begin{align}
    f_\text{mel} = 2595 \cdot \log_{10}{\left(1 + \frac{f_\text{Hz}}{700}\right)}.
\end{align}

Recently, transformer networks 
have been used with spectrograms~\cite{rs2010,mel} for general audio classification tasks~\cite{gong_2021_ssast, koutini_2021, msm_mae_niizumi}. 
Gong \etal proposed an audio spectrogram transformer,
which was further improved by augmenting self-supervision~\cite{gong_2021_ssast, koutini_2021, msm_mae_niizumi}.
Bartusiak \etal \cite{bartusiak_theasis} and Müller \etal \cite{in_the_wild} process spectrogram and mel-spectrogram using transformers for synthetic speech detection on ASVspoof2019 dataset. 
These approaches either process the spectrogram first using \glspl{cnn} and then use a transformer, or they process all regions of spectrogram together.
Koutini~\etal showed that processing spectrogram in patches can lead to higher performance in general audio classification tasks such as speaker recognition and environment sound classification.
Gong~\etal referred to frames as the set of all the patches in the spectrogram that correspond to the same temporal location and showed that frame representation leads to better performance on speech processing tasks than using patch representation.
Motivated by the promising performance of frame representation in general audio classification~\cite{gong_2021_ssast, koutini_2021} and speech attribution tasks~\cite{amit_ei_paper, assd_2023, acm_kratika_2023}, we propose \methodfullname.
Our proposed method processes mel-spectrogram in patches and then combines the representation from all the patches corresponding to same temporal location to get frame representation.

\begin{figure*}[!ht]
    \centering
    \includegraphics[width=.9\linewidth]{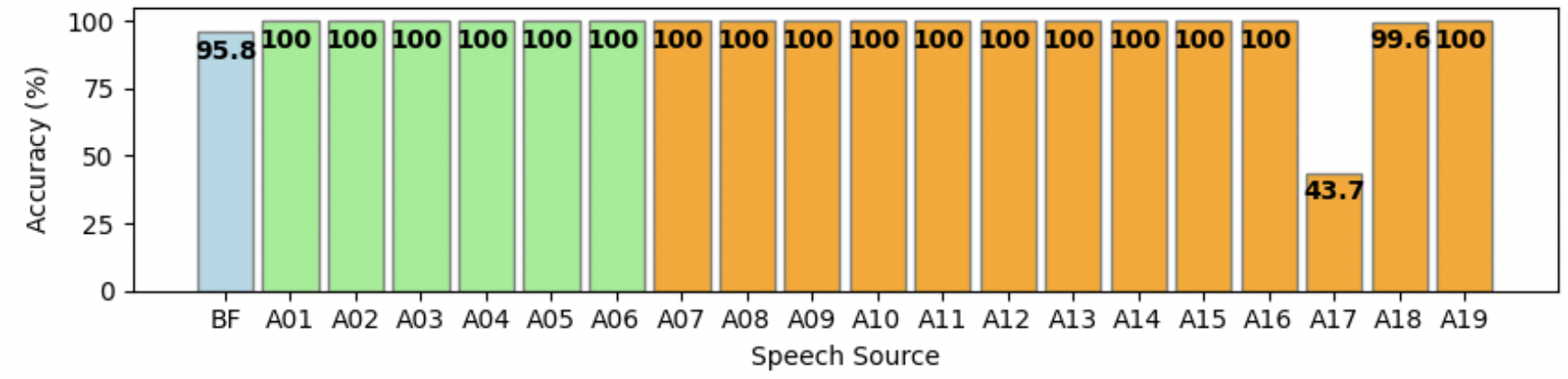}
    \caption{Detection accuracy of \methodname~on bona fide speech signals (blue), and synthesizers A01-A06 present in validation set D$_{dev}$ (green), and synthesizers A07-A19 present in the evaluation set D$_{eval}$ (orange) of the ASVspoof2019 dataset. }
    \label{fig:individual-accuracy}
    \label{fig:individual-accuracy}
\end{figure*}

%% file: part-3-method.tex
\section{Proposed Approach}\label{sec:method}
We propose \methodfullname~for detecting synthetic speech signals.
The block diagram of our proposed \methodname~is shown in \cref{fig:block-diagram}. 
Let $\Vec{x}$ be the time domain speech signal, we compute the magnitude of its Short Time Fourier Transform using a Hanning window of size 25 ms with a shift of 10 ms to obtain the mel-spectrogram~\cite{rs2010}.
Stevens \etal showed that the mel-scale correlates better with the human auditory system w.r.t to the frequency content than the Hz scale~\cite{mel}.
Mel-spectrograms have been used in general audio classification tasks~\cite{gong_2021_ssast, koutini_2021, msm_mae_niizumi} and we use them in~\methodname.
Following previous work~\cite{msm_mae_niizumi}, we consider $80$ frequency bins for creating the mel-spectrogram.
The size of the input speech signal is fixed to $5.12$ seconds, resulting in a mel-spectrogram of size $80\times 512$.
If the speech signal is less than $5.12$ seconds, we repeat the speech signal.
We divide the mel-spectrogram into patches of size $16\times16$.
We obtain $N=160$ patches. 
Using a linear layer, we convert each patch to a 768-dimensional vector representation $\Vec{e}_i, \; i \in \{1,2,..., N\}$ (see \cref{fig:block-diagram}).
To each patch representation (\ie $\Vec{e}_i$), we add a 768-dimensional vector $\Vec{p}_i$ corresponding to the position of the patch.

The resultant 768-dimensional representation for each patch (\ie $\Vec{e}_i + \Vec{p}_i$) is processed by a transformer encoder to get $\Vec{o}_i, \; i \in \{1,2,...,N\}$ (\cref{fig:block-diagram}). 
The architecture of the transformer encoder
is similar to the masked transformer encoder proposed for image classification~\cite{he2022masked} and for general audio classification~\cite{msm_mae_niizumi}.
The transformer encoder has a depth of 12 layers and uses 12 attention heads to obtain the patch representation. 
To avoid training the transformer from scratch we initialized our transformer encoder with pre-trained weights from~\cite{msm_mae_niizumi}.
Existing methods using transformer and spectrogram for synthetic speech detection mainly process all the regions of the spectrogram together and do not process patch representations \cite{bartusiak_theasis}. 
Contrary to existing methods, \methodname~uses patch representations. 
It rearranges patch representations to get frame representations.
The rearrangement is done by concatenating all the patch representations corresponding to same temporal position as shown in \cref{fig:block-diagram}.
This results in a frame representation (\ie $\Vec{f}_i, \; i \in \{1,2,..., L\}$, and $L = 32$)
(\cref{fig:block-diagram}).
We use the frame based representation because it has shown good performance in speech tasks such as synthetic speech attribution~\cite{amit_ei_paper} and general speech recognition tasks such as speaker identification~\cite{gong_2021_ssast}.
Also, we believe this makes \methodname~less sensitive to noise in one patch.

Finally, using a mean pooling layer we estimate the mean of all $\Vec{f_i}, \; i \in \{1,2,..., L\}$ to get $\Vec{f}$ as shown in \cref{fig:block-diagram}.
We process the mean frame representation $\Vec{f}$ using a \gls{mlp} network that consists of two linear layers separated by ReLU activation and a sigmoid activation at the last layer.
The \gls{mlp} network (\cref{fig:block-diagram}) provides two probabilities as outputs: $p_1$ and $p_2$ representing the probability of the speech signal being bona fide and synthetic, respectively.
To make the final decision on the input speech, we compare the two probabilities.
If the probability of the speech signal being bona fide is higher w.r.t it being synthetic (\ie ${p_1}>{p_2}$), we classify it as bona fide. 
Otherwise, we classify it as synthetic.

We use \gls{bce} loss for training because of its better performance in our experiments.
We trained for 50 epochs using a batch size of 256 and AdamW optimizer~\cite{loshchilov_2019} with an initial learning rate of $10^{-5}$ and a weight decay of $10^{-4}$.
We select the model weights which provide best accuracy on the validation set for evaluation.

%% file: part-4-results.tex
\glsresetall
\section{Experimental Results}\label{sec:experiment}
In this section, we describe the datasets used in each of our experiments, our experiments and discuss the results.
We use 
\gls{eer}~\cite{asvspoof19} as the performance metric
for all of our experiments. 
It is an official metric adopted by each of the datasets on which we evaluate our method.
We obtain \gls{eer} from \gls{roc} curve by finding the rate where \gls{fnr} and \gls{fpr} are equal.
Lower the \gls{eer}, the better the performance of the method. 
A perfect detector will have \gls{eer} of 0\% and a random classifier will have \gls{eer} of 50\%.

\subsection{Training and Validation Dataset}\label{sec:train_n_validate}
We use Logical Access (LA) part of the ASVspoof2019 dataset \cite{asvspoof19, asvdata_2019} for training and validation.
In total, there are 121,461 bona fide and synthetic speech signals divided with an approximate ratio of 1:1:3 into training set $D_{tr}$, validation set $D_{dev}$, and evaluation set $D_{eval}$~\cite{asvspoof19}.
Synthesized speech signals are generated using three diverse techniques - neural networks, vocoders, and waveform concatenation \cite{asvdata_2019}.
$D_{tr}$ and 
$D_{dev}$ contain approximately 89\% synthesized and 11\% bona fide speech signals.
The synthesized speech signals in 
$D_{tr}$ and 
$D_{dev}$ are generated from 6 different synthesizers (A01-A06).
Bona fide speech signals in each set are recorded from diverse human speakers that do not overlap among the two sets.
All speech signals are monophonic and lossless encoded using \gls{flac} format.


\glsresetall

\subsection{Experiment 1: Detection }\label{sec:experiment_1}
In this experiment, we investigate the detection performance of \methodfullname~on ASVspoof2019 dataset \cite{asvspoof19} and 
compare the performance
with existing methods.
For evaluation, we use the evaluation set $D_{eval}$ of the ASVspoof2019 dataset~\cite{asvspoof19}.
Section~\ref{sec:experiment_1_dataset} provides details about the evaluation set $D_{eval}$ \cite{asvspoof19} of the ASVspoof2019 dataset and Section~\ref{sec:experiment_1_result} discusses our experimental results on detecting synthetic speech signals.

\subsubsection{\textbf{Evaluation Dataset (ASVspoof2019)}}\label{sec:experiment_1_dataset}
The evaluation set $D_{eval}$ contains approximately 7.4k bona fide speech signals and 63.9k synthetic speech signals.
The synthetic speech is generated from 13 different synthesizers (A07-A19). 
The synthesizers A16 and A19 have same underlying architectures as A04 and A06 synthesizers present in the training $D_{tr}$ and validation set $D_{dev}$ of the ASVspoof2019 dataset. 
The remaining 11 synthesizers in the evaluation set $D_{eval}$ are unknown synthesizers \ie they have different underlying architectures 
than any synthesizers present in $D_{dev}$ or $D_{tr}$.
Overall, out of 63.9k synthesized speech signals in $D_{eval}$, 61.5k are generated from unknown synthesizers.
The bona fide speech signals in $D_{eval}$ are recorded from human speakers that do not overlap with the speakers in the 
$D_{tr}$ and 
$D_{dev}$.
Similar to $D_{tr}$ and $D_{dev}$, all speech signals in $D_{eval}$ are monophonic and lossless encoded using the \gls{flac} format.

\subsubsection{\textbf{Detection Performance}}\label{sec:experiment_1_result}
First, we report the detection performance of~\methodname~on the union of the $D_{dev}$ and the $D_{eval}$ sets.
It contains a total of 19 synthesizers (A01-A19). Synthesizers A01-A06 are present in 
$D_{dev}$ 
while A07-A19 are present in 
$D_{eval}$. 
\cref{fig:individual-accuracy} shows the performance of~\methodname~on different synthesizers and bona fide speech.
Our results show that~\methodname~has perfect detection accuracy for synthetic speech from known synthesizers that were also present in the training set (A01-A06). 
\methodname~even generalizes to unknown synthesizers and has almost perfect detection accuracy for synthetic speech from all of them except the $A17$ synthesizer.
$A17$ is also found as one of the most challenging class in the ASVspoof2019 Challenge result~\cite{asvspoof_2021}.
A possible reason for poor performance on $A17$ 
could be small duration of speech signal to analyze.
On average, signals from 
$A17$ have 26\% silence region in the start of the signal.
This is significantly higher than typical 8\% silence region in the start in other unknown synthesizers \eg $A14$.
Note that on average both of them have approximately same total duration of 3.4 seconds.
\begin{table}[!t]
    \centering
    \caption{Performance of Experiment 1 where PS3DT is compared with 14 spectrogram based methods and 2 baselines for synthetic speech detection on ASVspoof2019 dataset.}
    \input{tables/exp_1_performance_acc.tex}
    \label{tab:exp1-performance-aac}
\end{table}

Second, we report the \gls{eer} of \methodname~and compare it with that of 16 existing synthetic speech detection methods on the ASVspoof2019 dataset.
The proposed \methodfullname~uses mel-spectrogram to detect synthetic speech. 
Therefore, for comparison, we only include methods that either use spectrogram or mel-spectrogram or features obtained from spectrogram ($S01$- $S14$).
We also include the two baseline methods proposed and provided in the ASVspoof2019 Challenge ($B01$ and $B02$).
The two baseline comparison methods $B01$ and $B02$ process hand-crafted features like \gls{cqccs}, and \gls{lfccs} using a \gls{gmm}.
The remaining methods that we use for comparison convert the time domain signal to a spectrogram by either using \gls{stft} or using \gls{cqt}.
We refer to the spectrogram obtained using \gls{stft} and \gls{cqt} as Spectrogram and Spectrogram$_{CQT}$, respectively in \cref{tab:exp1-performance-aac}. 
Both of them are 2D spectral representations obtained from the time-domain speech signal.
The Spectrogram is obtained applying the Fourier transform to short overlapping segments of the signal and is computationally efficient. 
The Spectrogram$_{CQT}$ is obtained by decomposing the signal into a series of sinusoidal components that are logarithmically spaced in frequency. 
It is more computationally demanding than the Spectrogram but the frequency resolution is not dependent on the length of segment as it does not process the signal segment by segment.
Müller~\etal \cite{in_the_wild} showed that for a fixed neural network as in $S07$ to $S13$ using Spectrogram$_{CQT}$ results in better performance. 
Therefore, for comparison we only include versions of $S07$ to $S13$ that use Spectrogram$_{CQT}$ to detect synthetic speech.

Some methods compute logarithm of the spectrogram (shown as Log-Spectrogram in Table~\ref{tab:exp1-performance-aac}) or convert the spectrogram to mel-scale (shown as Mel-Spectrogram in Table~\ref{tab:exp1-performance-aac}). 
Methods $S05$ and $S06$ further process the Spectrogram$_{CQT}$ or inverted Spectrogram$_{CQT}$ using octave subbanding and \gls{dct} to obtain \gls{cqost} and \gls{icqcc} features, respectively.
Method $S04$ uses multiple features together, it uses both Spectrogram and \gls{cqt} \cite{rs2010} features to detect synthetic speech. We refer to it as Spectrogram$+CQT$ in \cref{tab:exp1-performance-aac}.
Each of the $S01$ to $S14$ methods use a neural network to process the spectrogram or its derivatives.
Similar to our proposed \methodname, methods $S07$ and $S14$ use transformer neural networks to detect synthetic speech.
$S14$ is the best performing transformer neural network based method proposed in \cite{bartusiak_theasis}.
We select it as a representative of all the transformer based methods proposed in \cite{bartusiak_theasis}. 
$S14$ is first trained using self-supervision on a large audio dataset \ie Audioset \cite{audioset}. 
Then using supervised learning, it is finetuned on ASVspoof2019 dataset \cite{asvspoof19} to detect synthetic speech.
Methods $S01$ and $S04$ are proposed in \cite{spec_vgg_sincnet_28} and they use VGG neural network~\cite{vgg_first_paper} and SincNet~\cite{spec_vgg_sincnet_28}.
Details about the neural networks used in $S02$, and $S07$ to $S13$ can be found in \cite{in_the_wild}.
Method $S03$ \cite{spec_cnn_telefor} uses a 3-layer \gls{cnn}.
Methods $S05$~\cite{tifs_cqost_da} and $S06$ \cite{spec_iqcc} use \gls{dnn} to detect synthetic speech.

We compare the performance of these 16 methods with~\methodname.
All 16 methods are trained, validated, and tested on the original $D_{train}$, $D_{dev}$, and $D_{eval}$ sets of the ASVspoof2019 dataset~\cite{asvspoof19}.
The results of this experiment are shown in~\cref{tab:exp1-performance-aac}.
\methodname~has around 3.5 percentage points and 5 percentage points improvement over the baseline methods $B01$ and $B02$, respectively. 
It has more than 5 percentage points improvement in \gls{eer} from spectrogram methods $S02$ and $S03$ using VGG and \gls{cnn} neural networks, respectively.
Compared to methods $S07$ and $S14$, which process Spectrogram$_{CQT}$ and Mel-Spectrogram using transformer neural networks, the improvement is around 3 percentage points and 0.7 percentage points, respectively.
From performance of $S14$, we can also note that using Mel-Spectrogram and large scale transformer such as \gls{passt} \cite{koutini_2021} results in better performance than even methods such as $S13$ and $S04$, which use fusion of neural networks for processing spectrogram.
Overall 
\methodname~has highest performance among the 14 different spectrogram based methods and the 2 baselines on ASVspoof2019 dataset.

\subsection{\textbf{Experiment 2 - Generalization}}\label{sec:experiment_2}
Existing work reports that speech signals from unknown synthesizers in ASVspoof2019 have similar silence distributions \cite{silence_gold}.
Some methods overfit on ASVspoof2019 dataset and use silence distribution instead of using features related to spoofing to detect synthetic speech \cite{silence_gold}.
Such methods can have high performance on detecting synthetic speech from unknown synthesizers in the ASVspoof2019 dataset. 
However, these methods are not good, as a simple attack such as removing silence can drastically drop their performance.
Müller~\etal proposed In-the-Wild dataset \cite{in_the_wild} to better investigate the generalization performance of methods trained on ASVspoof2019 dataset \cite{asvspoof19}.
In Experiment 2, we use this dataset to estimate the generalization performance of \methodname.
Note that for this experiment, we do not retrain \methodname~or any existing methods. 
We just evaluate cross dataset performance on the In-the-Wild dataset \cite{in_the_wild}.
We compared the generalization performance of \methodname~with top 8 existing detection methods ($S07$ to $S14$) discussed in Experiment 1 (\cref{tab:exp1-performance-aac}).
The details about In-the-Wild dataset are provided in Section~\ref{sec:experiment_2_generalization_dataset} and Section~\ref{sec:experiment_2_result} describes the \gls{eer} performance of \methodname~and existing methods on In-the-Wild dataset.

\subsubsection{\textbf{Evaluation Dataset (In-the-Wild)}}\label{sec:experiment_2_generalization_dataset}
The In-the-Wild dataset \cite{in_the_wild} contains approximately 
17 hours of high-quality synthetic speech impersonating 58 English-speaking politicians and celebrities and 21 hours of bona fide speech for the same politicians and celebrities.
They are downloaded from social networks and popular video sharing platforms.
The bona fide and synthetic speech are from diverse speakers, having different language accents, styles, and emotions.
The bona fide speech and the corresponding synthetic speech for a speaker have similar background noise, emotions, and duration. 
Each 
speech signal has an average length of 4.3 seconds and is transcoded to wav file format after downloading.
Each speech signal is converted to 8 bits and downsampled to 16 kHz.
\begin{table}[!t]
    \centering
    \caption{Experiment 2 generalization performance (\gls{eer}\%) on In-the-Wild dataset for top 8 performing existing detection methods from Experiment 1, and  \methodname~trained on ASVspoof2019 dataset.}

\input{tables/exp_2_performance.tex}
    \label{tab:exp2-performance}
\end{table}
\subsubsection{\textbf{Generalization Performance}}\label{sec:experiment_2_result}

The generalization performance of \methodname~and existing methods is shown in Table~\ref{tab:exp2-performance}.
The existing methods include the top 8 performing existing methods ($S07$ to $S14$) 
reported in Experiment 1.
Each of the methods is trained only on the training set of ASVspoof2019 dataset.
An \gls{eer} lower than 50\% is better than that of a random classifier and shows that the method generalizes to detect synthetic speech in the wild.
From Table~\ref{tab:exp2-performance}, we can see that methods $S08$, $S09$, $S10$, $S12$, $S13$ have performance worse than a random classifier. 
This maybe due to overfitting on the ASVspoof2019 dataset.
Transformer based methods $S07$ and $S14$ have better generalization performance.
We also notice that the method $S14$ which uses a self-supervised pretrained transformer, has the best performance after our proposed \methodname.
Overall, we can see that \methodname~generalizes better than all the existing spectrogram based methods for synthetic speech detection.
Müller~\etal in ~\cite{in_the_wild} also discuss the generalization performance of more than 55 synthetic speech detection methods trained on ASVspoof2019 dataset on In-the-Wild dataset.
Some of these detectors do not use spectrogram. 
The best performing method uses time domain speech signals and has generalization \gls{eer} of 33.94\% which improves marginally and becomes 33.10\% when the method is also trained on the validation set of the ASVspoof2019 dataset.
\methodname~has \gls{eer} of 29.72\% which is around 3\% percentage points better than the best performing method even when \methodname~does not use validation set for training.
Hence, our method has better generalization performance on In-the-Wild dataset than even other existing methods that do not necessarily use spectrogram for detecting synthetic speech.

\subsection{{Experiment 3 - Telephone Quality Robustness}}\label{sec:experiment_3_telephone}
In this experiment, we investigate the performance of~\methodname~for
authenticating speech signals over telephone channels.
We use evaluation set of Logical Access (LA) part of ASVspoof2021 dataset \cite{asvspoof_2021} for evaluation.
Section~\ref{sec:experiment_3_dataset} describes the dataset and Section~\ref{sec:experiment_3_result} discusses our results.

\subsubsection{\textbf{Evaluation Dataset (ASVspoof2021 LA)}}\label{sec:experiment_3_dataset}
We use the evaluation set of the Logical Access (LA) part of ASVspoof2021 dataset \cite{asvspoof_2021} for this experiment.
It contains 148K bona fide and synthetic speech signals.
The dataset is created by transmitting each speech signal in the evaluation set of the ASVspoof2019 dataset that we used in Experiment 1 over real telephone systems.
Two different systems were used for transmission namely, voice-over-internet-protocol (VoIP) system and a public switched telephone network (PSTN).
Each of the speech signals is transmitted using 7 different conditions. 
The first condition is a reference condition that does not process or transmit any speech signal over telephone and is identical to the ASVspoof2019 evaluation set~\cite{asvspoof19}. 
The remaining 6 conditions use 6 different commonly used telephone codecs for transmission. 
This include both legacy codecs such as a-law and G.722, and modern codecs such as OPUS and GSM. 
Common sampling rates used by telephone channels such as 8KHz and 16KHz were used for transmission.
The distance between the two endpoints of the telephone channel is significantly large \eg one endpoint is hosted in France and the other endpoint is either in Italy or Singapore.
More details about the dataset can be found in ~\cite{asvspoof_2021}.

\begin{table}[!t]
    \centering
    \caption{Experiment 3 robustness performance (\gls{eer}\%) for detecting synthetic speech over telephone channels using the ASVspoof2021 dataset for 4 baselines, top performing existing spectrogram method and proposed \methodname. The methods are trained on ASVspoof2019 dataset.}
    \input{tables/exp_3_asvspoof2021.tex}
    \label{tab:experiment_3_telephone}
\end{table}
\subsubsection{\textbf{Telephone Quality Robustness Performance}}\label{sec:experiment_3_result}
For investigating robustness of \methodname~over telephone channels, we use 
its version
trained on $D_{tr}$ set of ASVspoof2019 dataset (refer \cref{sec:train_n_validate}) and evaluated its performance on the evaluation set of ASVspoof2021 LA described in Section~\ref{sec:experiment_3_dataset}.
Table~\ref{tab:experiment_3_telephone} summarizes the performance of \methodname~and its comparison with that of four baselines provided in the ASVspoof2021 Challenge. 
The two baselines are the same as used in ASVspoof2019 \ie $B01$ and $B02$ in \cref{tab:exp1-performance-aac}.
The other two baselines $B03$ and $B04$ do not use spectrogram and were top performing submissions made to the ASVspoof2019 Challenge.
$B03$ 
processes \gls{lfccs} using a LCNN neural network and 
$B04$ 
uses time domain speech signals and processes them using \gls{dnn} \cite{asvspoof_2021}.
We also implemented $S14$ that 
showed best detection and generalization performance among all the existing methods in Experiment 1 and Experiment 2, respectively and evaluated its robustness performance.
In Table~\ref{tab:experiment_3_telephone},
it can be seen that
among all the baselines ($B01$ to $B04$) the method $B03$ has best robustness performance (\gls{eer} of 9.26\%). 
$S14$ has an \gls{eer} of 8.90\%. 
The results depict that processing mel-spectrogram using transformer has better robustness than baseline methods.
Overall, \methodname~with an \gls{eer} 8.29\% performs better than all the baselines $B01$ to $B04$ 
and 
$S14$.
Therefore, \methodname~has high robustness over telephone channels and can detect synthetic speech used over telephone to fool \gls{asv} systems or to impersonate a person.
\begin{table}[!b]
    \centering
    \caption{Experiment 4 robustness performance (\gls{eer}\%) of \methodname~for detecting compressed synthetic speech on each of the nine conditions in the ASVspoof2021 DF dataset. \methodname~was trained on ASVspoof2019 dataset. 
    }
    \input{tables/exp_4_asvspoof2021_df.tex}
    \label{tab:experiment_4_compression}
\end{table}

\subsection{Experiment 4: Robustness to Compression}\label{sec:experiment_4_compression}
In this experiment, we investigate the performance of~\methodname~when deployed in practical scenarios such as authenticating speech signals uploaded on social platforms.
We use the DeepFake (DF) part of the ASVspoof2021 dataset \cite{asvspoof_2021} for our evaluation.
Section~\ref{sec:experiment_4_dataset} provides more details about the dataset and Section~\ref{sec:experiment_4_result} describes our results.

\subsubsection{\textbf{Evaluation Dataset (ASVspoof2021 DF)}}\label{sec:experiment_4_dataset}
We use the DeepFake (DF) part of the ASVspoof2021 dataset for this experiment \cite{asvspoof_2021}.
It contains approximately 612K bona fide and synthetic speech signals.
The synthetic speech is generated from more than 100 different synthesizers.
The speech signals are lossy encoded using different standards and using both high and low variable data/bit rates.
The uncompressed speech samples are obtained by a union of speech samples from three datasets, namely the evaluation set of ASVspoof2019 LA dataset \cite{asvspoof19}, the 2018 and 2020 Voice Conversion Challenge (VCC) datasets \cite{vcc_2018, vcc_2020}.
Depending on the encoding standard and data rate used for compression, there are nine partitions in the dataset referred to as nine different conditions ($DF-C1$ to $DF-C9$).
The uncompressed partition of speech set is referred as $DF-C1$ in Table~\ref{tab:experiment_4_compression}.
The other eight partitions are created by compressing $DF-C1$ partition using different standards and data rates.
Table~\ref{tab:experiment_4_compression} shows compression standards and variable bit rates used in each of the conditions.
Notice that conditions $DF-C2$ to $DF-C7$ are encoding uncompressed speech to compressed speech. 
While conditions $DF-C8$ and $DF-C9$ first encode the uncompressed speech signal to compressed speech using a particular compression standard and then further transcode compressed speech signal to \gls{aac} \cite{mpeg4_book} compression standard. 
Together, the conditions $DF-C1$ to $DF-C9$ represent diverse compression scenarios which occur when synthetic speech is uploaded on different social platforms.
For instance, YouTube typically uses \gls{aac} \cite{youtube_aac}.

\begin{table}[!t]
    \centering
    \caption{Experiment 4 robustness performance (\gls{eer}\%) of 4 baselines, top performing existing spectrogram method and proposed \methodname~for detecting compressed synthetic speech on ASVspoof2021 DF dataset. The methods are trained on ASVspoof2019 dataset.
    }
    \input{tables/exp_4_asvspoof2021_compression.tex}
    \label{tab:experiment_4_compression_2}
\end{table}
\subsubsection{\textbf{Robustness to Compression Performance}}\label{sec:experiment_4_result}
For investigating robustness of~\methodname~to compression, we use the version of~\methodname~trained on $D_{tr}$ set of ASVspoof2019 dataset \cite{asvspoof19} and evaluated its performance on the ASVspoof2021 DF dataset (described in detail in Section~\ref{sec:experiment_4_dataset}).
First, in Table~\ref{tab:experiment_4_compression} we show performance of \methodname~on each of the nine partitions of the dataset.
As expected the performance of \methodname~is good on uncompressed conditions \ie $DF-C1$.
Also, out of MP3, OGG, and AAC compression standards \cite{mpeg4_book, isomp3} \methodname~is most robust to AAC compression that is used by several social platforms such as YouTube and Twitter \cite{youtube_aac}.
Surprisingly, the performance is better on \gls{aac} compressed signal than on uncompressed speech signal. 
A possible reason could be that
\gls{aac} compression is one of the most efficient and recent compression techniques \cite{mpeg4_book}.
The spectral processing in it includes blocks to process synthetic speech differently that might be making spoofing artifacts more evident in the spectrogram \cite{mpeg4_book}. 
Some previous work show that \gls{aac} compression has information in the encoding bit stream that can even help to detect synthetic speech \cite{assd_2023}. 
\methodname~is least robust to MP3 compression standard.
Also, as expected lower data rates during compression will distort speech signal more and hence \methodname~has lower performance for all MP3, OGG, and AAC compression standards at lower data rates as compared to that for higher data rates.
Further, the performance is lower on transcoding where multiple compressions are done than on single compression \ie encoding.
Out of transcoding conditions \ie $DF-C8$ and $DF-C9$, the performance is lower if the first compression is done using MP3 standard (\ie in $DF-C8$) as compared to that if the first compression is done using OGG standard.
A possible reason could be that single low rate MP3 compression reduces performance of \methodname~more significantly as observed by lowest performance in $DF-C2$ condition.
We also found accuracy of \methodname~for each condition in \cref{tab:experiment_4_compression}.
The accuracy is the ratio of the total correct classifications to the total number of classifications.
There are around 68K bona fide and synthetic speech signals in each condition out of which 65K speech signals are synthetic. 
For each condition, \methodname~has accuracy of 97\% or higher.

Second, we compare the robustness of \methodname~to detect compressed synthetic speech with other methods.
Table~\ref{tab:experiment_4_compression_2} compares the robustness to compression performance of $S14$ and the same 4 baselines used in Experiment 3 and provided in ASVspoof2021 Challenge\cite{asvspoof_2021}.
Among all the baselines ($B01$ to $B04$), the method $B04$ has best robustness performance (\gls{eer} of 22.38\%). 
$S14$ has an \gls{eer} of 24.37\%.
Comparing these results with robustness to detect synthetic speech over telephone in \cref{tab:experiment_3_telephone}, we observe that if one method is more robust than other method in detecting synthetic speech over telephone then it may not necessarily be more robust in detecting compressed synthetic speech. 
For instance, baseline methods $B03$ and method $S14$ are more robust than baseline $B04$ in \cref{tab:experiment_3_telephone}, however, their performance is lower than that of $B04$ in \cref{tab:experiment_4_compression_2}.
\methodname~performs consistently better than all the baseline methods and method $S14$ in both \cref{tab:experiment_3_telephone} and \cref{tab:experiment_4_compression_2}.

Overall, \methodname~performs significantly better than all the baselines $B01$ to $B04$ and method $S14$ with \gls{eer} 16.61\%.
We also compared robustness of \methodname~with 29 methods other than the baseline methods submitted to ASVspoof2021 DeepFake Challenge \cite{asvspoof_2021}.
These methods do not necessarily use spectrogram but use other approaches \eg fusion of several features to detect synthetic speech. 
More details about them can be found in \cite{asvspoof_2021}.
The least performing method has an \gls{eer} of 29.75\% while the best performing method has \gls{eer} of 15.64\%. 
\methodname~has third best performance with an \gls{eer} of 16.61\% as compared to all the 29 methods reported in \cite{asvspoof_2021}.
Therefore, \methodname~has high robustness to compression and hence can detect synthetic speech shared on social platforms to spread misinformation or impersonate a person.

%% file: tables/exp_1_performance_acc.tex

\begin{tabular}{@{\extracolsep{-4pt}}lcccc}
    \toprule
     Method Name & Feature & Network & EER \\\midrule
     $B01$ & CQCC  & GMM & 8.09\% \\
     $B02$ & LFCC  & GMM & 9.57\% \\
     $S01$ & Spectrogram & VGG & 10.52\%\\
     $S02$ & Log-Spectrogram & MesoInception & 10.02\%\\
     $S03$ & Spectrogram & CNN & 9.57\%\\
     $S04$ & Spectrogram+CQT & VGG+SincNet & 8.01\% \\
     $S05$ & Spectrogram$_{CQOST}$ & DNN & 8.04\% \\
     $S06$ & Spectrogram$_{ICQCC}$ & DNN & 7.70\% \\
     $S07$ & Spectrogram$_{CQT}$ & Transformer & 7.50\% \\
     $S08$ & Spectrogram$_{CQT}$ & MesoNet & 7.42\% \\
     $S09$ & Spectrogram$_{CQT}$ & LSTM &  7.16\%\\
     $S10$ & Spectrogram$_{CQT}$ & LCNN-Attention & 6.76\% \\
     $S11$ & Spectrogram$_{CQT}$ & ResNet18 & 6.55\%\\
     $S12$ & Spectrogram$_{CQT}$ & LCNN & 6.35\% \\
     $S13$ & Spectrogram$_{CQT}$ & LCNN+LSTM & 6.23\% \\
     $S14$ & Mel-Spectrogram & PaSST & 5.26\% \\
     \textbf{\methodname} & Mel-Spectrogram & Patched Transformer & \textbf{4.54\%} \\
     \bottomrule
     \end{tabular}

%% file: tables/exp_2_performance.tex
\begin{tabular}{@{\extracolsep{-4pt}}lcccc}
    \toprule
     Method Name & Feature & Network & EER \\\midrule
     $S07$ & Spectrogram$_{CQT}$ & Transformer & 43.78\% \\
     $S08$ & Spectrogram$_{CQT}$ & MesoNet & 54.54\% \\
     $S09$ & Spectrogram$_{CQT}$ & LSTM &  53.71\%\\
     $S10$ & Spectrogram$_{CQT}$ & LCNN-Attention & 66.68\% \\
     $S11$ & Spectrogram$_{CQT}$ & ResNet18 & 49.76\%\\
     $S12$ & Spectrogram$_{CQT}$ & LCNN & 65.56\% \\
     $S13$ & Spectrogram$_{CQT}$ & LCNN+LSTM & 61.50\% \\
     $S14$ & Mel-Spectrogram & PaSST & 39.98\% \\
     \textbf{\methodname} & Mel-Spectrogram & Patched Transformer & \textbf{29.72\% } \\
     \bottomrule
     \end{tabular}

%% file: tables/exp_3_asvspoof2021.tex
\begin{tabular}{@{\extracolsep{-4pt}}lcccc}
    \toprule
     Method Name & Feature & Network & EER \\\midrule
     $B01$ & CQCC  & GMM & 15.62\% \\
     $B02$ & LFCC  & GMM & 19.30\% \\
     $B03$ & LFCC & LCNN & 9.26\% \\
     $B04$ & time-domain & DNN & 9.50\% \\
     S14 & Mel-Spectrogram & PaSST & 8.90\% \\
     \textbf{\methodname} & Mel-Spectrogram & Patched Transformer & \textbf{8.29\%} \\
     \bottomrule
     \end{tabular}

%% file: tables/exp_4_asvspoof2021_df.tex
\begin{tabular}{@{\extracolsep{-4pt}}lcccc}
    \toprule
     Condition & Compression & Data rate & EER \\\midrule
     $DF-C1$ & No compression  & 256 kbps & 14.80\% \\
     $DF-C2$ & Low MP3  & 80-120 kbps & 21.08\% \\
     $DF-C3$ & High MP3 & 220-260 kbps  & 20.02\% \\
     $DF-C4$ & Low AAC & 20-32 kbps & 14.70\% \\
     $DF-C5$ & High AAC & 96-112 kbps & 14.11\% \\
     $DF-C6$ & Low OGG & 80-96 kbps & 16.22\% \\
     $DF-C7$ & High OGG & 256-320 kbps & 14.38\% \\
     $DF-C8$ & MP3 $\rightarrow$ AAC  & 80-120 kbps $\rightarrow$ 96-112 kbps & 21.90\% \\
     $DF-C9$ & OGG $\rightarrow$ AAC & 80-96 kbps $\rightarrow$ 96-112 kbps & 15.36\% \\
     \bottomrule
     \end{tabular}

%% file: tables/exp_4_asvspoof2021_compression.tex
\begin{tabular}{@{\extracolsep{-4pt}}lcccc}
    \toprule
     Method Name & Feature & Network & EER \\\midrule
     $B01$ & CQCC  & GMM & 25.56\% \\
     $B02$ & LFCC  & GMM & 25.25\% \\
     $B03$ & LFCC & LCNN & 23.48\%\\
     $B04$ & time-domain & DNN & 22.38\% \\
     S14 & Mel-Spectrogram & PaSST & 24.37 \% \\
     \textbf{\methodname} & Mel-Spectrogram & Patched Transformer & \textbf{16.61\%}\\
     \bottomrule
     \end{tabular}

%% file: part-5-conclusion.tex
\section{Conclusion}\label{sec:conclusion}
In this paper we proposed \methodfullname~for
synthetic speech detection.
\methodname~has perfect detection accuracy for 6 known and 10 out of 11 unknown speech synthesizers.
\methodname~does better than previous methods which use either spectrogram or its derivatives for synthetic speech detection.
We show that \methodname~generalizes better than existing methods and works better in practical scenarios such as detecting synthetic speech shared on social platforms and telephone quality synthetic speech.
In future, we will work on localization of partially synthetic speech. 
We also plan to explore a multimodal approach that uses both spectral representation \eg mel-spectrogram, and time-domain speech signal for detection.


%% file: main.bbl
\begin{thebibliography}{10}
\providecommand{\url}[1]{#1}
\csname url@samestyle\endcsname
\providecommand{\newblock}{\relax}
\providecommand{\bibinfo}[2]{#2}
\providecommand{\BIBentrySTDinterwordspacing}{\spaceskip=0pt\relax}
\providecommand{\BIBentryALTinterwordstretchfactor}{4}
\providecommand{\BIBentryALTinterwordspacing}{\spaceskip=\fontdimen2\font plus
\BIBentryALTinterwordstretchfactor\fontdimen3\font minus \fontdimen4\font\relax}
\providecommand{\BIBforeignlanguage}[2]{{%
\expandafter\ifx\csname l@#1\endcsname\relax
\typeout{** WARNING: IEEEtran.bst: No hyphenation pattern has been}%
\typeout{** loaded for the language `#1'. Using the pattern for}%
\typeout{** the default language instead.}%
\else
\language=\csname l@#1\endcsname
\fi
#2}}
\providecommand{\BIBdecl}{\relax}
\BIBdecl

\bibitem{asvspoof19}
M.~Todisco, X.~Wang, V.~Vestman, M.~Sahidullah, H.~Delgado, A.~Nautsch, J.~Yamagishi, N.~Evans, T.~Kinnunen, and K.~A. Lee, ``{ASVspoof 2019: Future Horizons in Spoofed and Fake Audio Detection},'' \emph{Proceedings of the Interspeech}, pp. 1008--1012, September 2019, {Graz, Austria}.

\bibitem{popov2021gradtts}
V.~Popov, I.~Vovk, V.~Gogoryan, T.~Sadekova, and M.~Kudinov, ``{Grad-TTS: A Diffusion Probabilistic Model for Text-to-Speech},'' \emph{Proceedings of the International Conference on Machine Learning}, vol. 139, pp. 8599--8608, July 2021, {Virtual}.

\bibitem{klatt2016}
D.~H. Klatt, ``{Review of Text‐to‐Speech Conversion for English},'' \emph{The Journal of the Acoustical Society of America}, vol.~82, no.~3, p. 737–793, May 1987.

\bibitem{fastspeech_2_iclr_2021}
Y.~Ren, C.~Hu, X.~Tan, T.~Qin, S.~Zhao, Z.~Zhao, and T.-Y. Liu, ``{FastSpeech 2: Fast and High-Quality End-to-End Text to Speech},'' \emph{Proceedings of the International Conference on Learning Representations}, pp. 1--15, May 2021, {virtual}.

\bibitem{wellsaid}
\BIBentryALTinterwordspacing
{WellSaid Labs, Inc. 2022}, ``{WELLSAID: AI Voice Over for Commercials},'' 2022. [Online]. Available: \url{https://wellsaidlabs.com/ai-voice-over}
\BIBentrySTDinterwordspacing

\bibitem{vall_e}
C.~Wang, S.~Chen, Y.~Wu, Z.~Zhang, L.~Zhou, S.~Liu, Z.~Chen, Y.~Liu, H.~Wang, J.~Li, L.~He, S.~Zhao, and F.~Wei, ``{Neural Codec Language Models are Zero-Shot Text to Speech Synthesizers},'' \emph{arXiv preprint}, January 2023.

\bibitem{allyn_2022}
B.~Allyn, ``{Deepfake Video of Zelenskyy Could be `Tip of the Iceberg' in Info War, Experts Warn},'' \url{https://www.npr.org/2022/03/16/1087062648/deepfake-video-zelenskyy-experts-war-manipulation-ukraine-russia}, March 2022.

\bibitem{smith_2021}
\BIBentryALTinterwordspacing
B.~Smith, ``{Goldman Sachs, Ozy Media and a \$40 Million Conference Call Gone Wrong},'' \emph{The New York Times}, September 2021. [Online]. Available: \url{https://www.nytimes.com/2021/09/26/business/media/ozy-media-goldman-sachs.html}
\BIBentrySTDinterwordspacing

\bibitem{bhagtani2022overview}
K.~Bhagtani, A.~K.~S. Yadav, E.~R. Bartusiak, Z.~Xiang, R.~Shao, S.~Baireddy, and E.~J. Delp, ``{An Overview of Recent Work in Multimedia Forensics},'' \emph{Proceedings of the IEEE Conference on Multimedia Information Processing and Retrieval}, pp. 324--329, August 2022, {Virtual}.

\bibitem{mpeg4_book}
J.~Herre and H.~Purnhagen, ``{General Audio Coding},'' in \emph{{The MPEG-4 Book}}, F.~C. Pereira and T.~Ebrahimi, Eds.\hskip 1em plus 0.5em minus 0.4em\relax Upper Saddle River, NJ, USA: Prentice Hall PTR, 2002, pp. 487--544.

\bibitem{youtube_aac}
\BIBentryALTinterwordspacing
{Google Inc.}, ``{YouTube Recommended Upload Encoding Settings},'' 2022. [Online]. Available: \url{https://support.google.com/youtube/answer/1722171}
\BIBentrySTDinterwordspacing

\bibitem{bartusiak_2021_asilomar}
E.~R. Bartusiak and E.~J. Delp, ``{Synthesized Speech Detection Using Convolutional Transformer-Based Spectrogram Analysis},'' \emph{Proceedings of the IEEE Asilomar Conference on Signals, Systems, and Computers}, pp. 1426--1430, October 2021, {Asilomar, CA}.

\bibitem{bartusiak_theasis}
E.~R. Bartusiak, ``{Machine Learning for Speech Forensics and Hypersonic Vehicle Applications},'' Ph.D. dissertation, Purdue University, West Lafayette, IN, 12 2022.

\bibitem{mel}
S.~S. Stevens, J.~Volkmann, and E.~B. Newman, ``{A Scale for the Measurement of the Psychological Magnitude Pitch},'' \emph{Journal of the Acoustical Society of America}, vol.~8, pp. 185--190, June 1937.

\bibitem{rs2010}
L.~Rabiner and R.~Schafer, \emph{{Theory and Applications of Digital Speech Processing}}, 1st~ed.\hskip 1em plus 0.5em minus 0.4em\relax USA: Prentice Hall Press, 2010.

\bibitem{in_the_wild}
N.~M. M{\"u}ller, P.~Czempin, F.~Dieckmann, A.~Froghyar, and K.~B{\"o}ttinger, ``Does audio deepfake detection generalize?'' \emph{Proceedings of the Interspeech}, September 2022, {Incheon, Korea}.

\bibitem{asvspoof_2021}
X.~Liu, X.~Wang, M.~Sahidullah, J.~Patino, H.~Delgado, T.~Kinnunen, M.~Todisco, J.~Yamagishi, N.~Evans, A.~Nautsch \emph{et~al.}, ``Asvspoof 2021: Towards spoofed and deepfake speech detection in the wild,'' \emph{arXiv preprint}, 2022.

\bibitem{li2021replay}
X.~Li, N.~Li, C.~Weng, X.~Liu, D.~Su, D.~Yu, and H.~Meng, ``{Replay and Synthetic Speech Detection with Res2Net Architecture},'' \emph{Proceedings of the IEEE International Conference on Acoustics, Speech and Signal Processing}, pp. 6354--6358, June 2021, {Toronto, Canada}.

\bibitem{cqccs}
M.~Todisco, H.~Delgado, and N.~Evans, ``{Constant Q Cepstral Coefficients: A Spoofing Countermeasure for Automatic Speaker Verification},'' \emph{Computer Speech \& Language}, vol.~45, pp. 516--535, September 2017.

\bibitem{mfccs}
M.~Sahidullah and G.~Saha, ``{Design, Analysis, and Experimental Evaluation of Block Based Transformation in MFCC Computation for Speaker Recognition},'' \emph{Speech Communication}, vol.~54, pp. 543--565, May 2012.

\bibitem{zakariah2017digital}
M.~Zakariah, M.~K. Khan, and H.~Malik, ``{Digital Multimedia Audio Forensics: Past, Present and Future},'' \emph{Multimedia Tools and Applications}, vol.~77, no.~1, pp. 1009--1040, January 2017.

\bibitem{he2016deep}
K.~He, X.~Zhang, S.~Ren, and J.~Sun, ``{Deep Residual Learning for Image Recognition},'' \emph{Proceedings of the IEEE Conference on Computer Vision and Pattern Recognition}, pp. 770--778, June 2016, {Las Vegas, NV}.

\bibitem{subramani2020learning}
N.~Subramani and D.~Rao, ``{Learning Efficient Representations for Fake Speech Detection},'' \emph{Proceedings of the AAAI Conference on Artificial Intelligence}, vol.~34, no.~04, pp. 5859--5866, April 2020, {New York, NY}.

\bibitem{long-mipr}
K.~Bhagtani, A.~K.~S. Yadav, E.~R. Bartusiak, Z.~Xiang, R.~Shao, S.~Baireddy, and E.~J. Delp, ``{An Overview of Recent Work in Media Forensics: Methods and Threats},'' \emph{arXiv preprint arXiv:2204.12067}, April 2022.

\bibitem{alan_acm_2023}
Z.~Xiang, A.~K.~S. Yadav, S.~Tubaro, P.~Bestagini, and E.~J. Delp, ``Extracting efficient spectrograms from mp3 compressed speech signals for synthetic speech detection,'' \emph{Proceedings of the ACM Workshop on Information Hiding and Multimedia Security}, p. 163–168, 2023, {Chicago, IL, USA}.

\bibitem{gong_2021_ssast}
Y.~Gong, C.-I. Lai, Y.-A. Chung, and J.~Glass, ``{SSAST: Self-Supervised Audio Spectrogram Transformer},'' \emph{Proceedings of the AAAI Conference on Artificial Intelligence}, vol.~36, no.~10, pp. 10\,699--10\,709, October 2022, {Virtual}.

\bibitem{koutini_2021}
K.~Koutini, J.~Schlüter, H.~Eghbal-zadeh, and G.~Widmer, ``{Efficient Training of Audio Transformers with Patchout},'' \emph{Proceedings of the Interspeech}, pp. 2753--2757, September 2022, {Incheon, Korea}.

\bibitem{msm_mae_niizumi}
D.~Niizumi, D.~Takeuchi, Y.~Ohishi, N.~Harada, and K.~Kashino, ``{Masked Spectrogram Modeling using Masked Autoencoders for Learning General-purpose Audio Representation},'' \emph{Proceedings of Machine Learning Research}, vol. 166, pp. 1--24, Dec 2022.

\bibitem{amit_ei_paper}
A.~K.~S. Yadav, E.~Bartusiak, K.~Bhagtani, and E.~J. Delp, ``Synthetic speech attribution using self supervised audio spectrogram transformer,'' \emph{Proceedings of the IS\&T Media Watermarking, Security, and Forensics Conference, Electronic Imaging Symposium}, January 2023, san Francisco, CA.

\bibitem{assd_2023}
A.~K. Singh~Yadav, Z.~Xiang, E.~R. Bartusiak, P.~Bestagini, S.~Tubaro, and E.~J. Delp, ``{ASSD: Synthetic Speech Detection in the AAC Compressed Domain},'' \emph{Proceedings of the IEEE International Conference on Acoustics, Speech, and Signal Processing}, pp. 1--5, June 2023, {Rhodes Island, Greece}.

\bibitem{acm_kratika_2023}
K.~Bhagtani, E.~R. Bartusiak, A.~K.~S. Yadav, P.~Bestagini, and E.~J. Delp, ``Synthesized speech attribution using the patchout spectrogram attribution transformer,'' \emph{Proceedings of the ACM Workshop on Information Hiding and Multimedia Security}, p. 157–162, June 2023, {Chicago, IL, USA}.

\bibitem{he2022masked}
K.~He, X.~Chen, S.~Xie, Y.~Li, P.~Doll{\'a}r, and R.~Girshick, ``Masked autoencoders are scalable vision learners,'' \emph{Proceedings of the IEEE/CVF Conference on Computer Vision and Pattern Recognition}, pp. 16\,000--16\,009, June 2022.

\bibitem{loshchilov_2019}
I.~Loshchilov and F.~Hutter, ``{Decoupled Weight Decay Regularization},'' \emph{Proceedings of the International Conference on Learning Representations}, May 2019, {New Orleans, LA}.

\bibitem{asvdata_2019}
\BIBentryALTinterwordspacing
J.~Yamagishi, M.~Todisco, M.~Sahidullah, H.~Delgado, X.~Wang, N.~Evans, T.~Kinnunen, K.~Lee, V.~Vestman, and A.~Nautsch, ``{ASVspoof 2019: The 3rd Automatic Speaker Verification Spoofing and Countermeasures Challenge database},'' March 2019. [Online]. Available: \url{https://www.asvspoof.org/index2019.html}
\BIBentrySTDinterwordspacing

\bibitem{audioset}
J.~F. Gemmeke, D.~P. Ellis, D.~Freedman, A.~Jansen, W.~Lawrence, R.~C. Moore, M.~Plakal, and M.~Ritter, ``{Audio set: An Ontology and Human-labeled Dataset for Audio Events},'' \emph{Proceedings of the IEEE International Conference on Acoustics, Speech and Signal Processing}, March 2017, {New Orleans, LA}.

\bibitem{spec_vgg_sincnet_28}
H.~Zeinali, T.~Stafylakis, G.~Athanasopoulou, J.~Rohdin, I.~Gkinis, L.~Burget, and J.~{\v{C}}ernock{\`y}, ``Detecting spoofing attacks using vgg and sincnet: But-omilia submission to asvspoof 2019 challenge,'' \emph{Proceedings of the Interspeech}, pp. 1073--1077, {September} 2019, {Graz, Austria}.

\bibitem{vgg_first_paper}
K.~Simonyan and A.~Zisserman, ``{Very Deep Convolutional Networks for Large-Scale Image Recognition},'' \emph{arXiv preprint arXiv:1409.1556}, 2014.

\bibitem{spec_cnn_telefor}
T.~Nosek, S.~Suzić, B.~Papić, and N.~Jakovljević, ``{Synthesized Speech Detection Based on Spectrogram and Convolutional Neural Networks},'' \emph{Proceedings of the IEEE Telecommunications Forum}, pp. 1--4, November 2019, {Belgrade, Serbia}.

\bibitem{tifs_cqost_da}
J.~Yang, R.~K. Das, and H.~Li, ``{Significance of Subband Features for Synthetic Speech Detection},'' \emph{IEEE Transactions on Information Forensics and Security}, vol.~15, pp. 2160--2170, November 2020.

\bibitem{spec_iqcc}
J.~Yang and R.~K. Das, ``{Long-term high frequency features for synthetic speech detection},'' \emph{Digital Signal Processing}, vol.~97, p. 102622, February 2020.

\bibitem{silence_gold}
N.~M. M{\"u}ller, F.~Dieckmann, P.~Czempin, R.~Canals, K.~B{\"o}ttinger, and J.~Williams, ``Speech is silver, silence is golden: What do asvspoof-trained models really learn?'' \emph{arXiv preprint}, 2021.

\bibitem{vcc_2018}
J.~Lorenzo-Trueba, J.~Yamagishi, T.~Toda, D.~Saito, F.~Villavicencio, T.~Kinnunen, and Z.~Ling, ``The voice conversion challenge 2018: Promoting development of parallel and nonparallel methods,'' \emph{Proceedings of the Speaker and Language Recognition Workshop}, pp. 195--202, June 2018, {Les Sables d'Olonne, France}.

\bibitem{vcc_2020}
Z.~Yi, W.-C. Huang, X.~Tian, J.~Yamagishi, R.~K. Das, T.~Kinnunen, Z.-H. Ling, and T.~Toda, ``{Voice Conversion Challenge 2020 -- Intra-lingual semi-parallel and cross-lingual voice conversion},'' \emph{Proceedings of the Joint Workshop for the Blizzard Challenge and Voice Conversion Challenge 2020}, pp. 80--98, october 2020, {Shanghai, China}.

\bibitem{isomp3}
\BIBentryALTinterwordspacing
{International Organization for Standardization/International Electrotechnical Commission}, ``{ISO/IEC 13818-3:1995 Information technology - Generic Coding of Moving Pictures and Associated Audio Information - Part 3: Audio},'' 1995. [Online]. Available: \url{https://www.iso.org/standard/22991.html}
\BIBentrySTDinterwordspacing

\end{thebibliography}
